\documentclass[aps,showpacs,showkeys]{revtex4}
\usepackage{amsmath,amsthm,amssymb,epsfig,alltt}

\begin{document}

\def\a{\alpha}
\def\b{\beta}
\def\d{{\delta}}
\def\l{\lambda}
\def\e{\epsilon}
\def\p{\partial}
\def\m{\mu}
\def\n{\nu}
\def\t{\tau}
\def\th{\theta}
\def\s{\sigma}
\def\g{\gamma}
\def\o{\omega}
\def\r{\rho}
\def\half{\frac{1}{2}}
\def\hatt{{\hat t}}
\def\hatx{{\hat x}}
\def\hatp{{\hat p}}
\def\hatX{{\hat X}}
\def\hatY{{\hat Y}}
\def\hatP{{\hat P}}
\def\haty{{\hat y}}
\def\whatX{{\widehat{X}}}
\def\whata{{\widehat{\alpha}}}
\def\whatb{{\widehat{\beta}}}
\def\whatV{{\widehat{V}}}
\def\hatth{{\hat \theta}}
\def\hatta{{\hat \tau}}
\def\hatrh{{\hat \rho}}
\def\hatva{{\hat \varphi}}
\def\barx{{\bar x}}
\def\bary{{\bar y}}
\def\barz{{\bar z}}
\def\baro{{\bar \omega}}
\def\barpsi{{\bar \psi}}
\def\sp{\sigma^\prime}
\def\nn{\nonumber}
\def\cb{{\cal B}}
\def\2pap{2\pi\alpha^\prime}
\def\wideA{\widehat{A}}
\def\wideF{\widehat{F}}
\def\beq{\begin{eqnarray}}
 \def\eeq{\end{eqnarray}}
 \def\4pap{4\pi\a^\prime}
 \def\xp{{x^\prime}}
 \def\sp{{\s^\prime}}
 \def\ap{{\a^\prime}}
 \def\tp{{\t^\prime}}
 \def\zp{{z^\prime}}
 \def\xpp{x^{\prime\prime}}
 \def\xppp{x^{\prime\prime\prime}}
 \def\barxp{{\bar x}^\prime}
 \def\barxpp{{\bar x}^{\prime\prime}}
 \def\barxppp{{\bar x}^{\prime\prime\prime}}
 \def\barchi{{\bar \chi}}
 \def\baro{{\bar \omega}}
 \def\bpsi{{\bar \psi}}
 \def\barg{{\bar g}}
 \def\barz{{\bar z}}
 \def\bareta{{\bar \eta}}
 \def\ta{{\tilde \a}}
 \def\tb{{\tilde \b}}
 \def\tc{{\tilde c}}
 \def\tz{{\tilde z}}
 \def\tJ{{\tilde J}}
 \def\tpsi{\tilde{\psi}}
 \def\tal{{\tilde \alpha}}
 \def\tbe{{\tilde \beta}}
 \def\tga{{\tilde \gamma}}
 \def\tchi{{\tilde{\chi}}}
 \def\barth{{\bar \theta}}
 \def\bareta{{\bar \eta}}
 \def\barom{{\bar \omega}}
 \def\bole{{\boldsymbol \epsilon}}
 \def\bolth{{\boldsymbol \theta}}
 \def\bomega{{\boldsymbol \omega}}
 \def\bolmu{{\boldsymbol \mu}}
 \def\bola{{\boldsymbol \alpha}}
 \def\bolb{{\boldsymbol \beta}}
 \def\bolvarphi{{\boldsymbol \varphi}}
 \def\bolN{{\boldsymbol N}}

\setcounter{page}{1}
\title[]{Klein factors and Fermi-Bose Equivalence }

\author{Taejin Lee}
\affiliation{
Department of Physics, Kangwon National University, Chuncheon 200-701
Korea}

\email{taejin@kangwon.ac.kr}

\begin{abstract}
Generalizing the kink operator of the Heisenberg spin 1/2 model, we construct a set of Klein factors explicitly such that $(1+1)$ dimensional fermion theories with arbitrary number of species are mapped onto the corresponding boson theories with the same number of species and vice versa. The actions for the resultant theories do not possess any nontrivial Klein factor. With this set of Klein factors, we are also able to map the simple boundary states such as the Neumann and the Dirichlet boundary states, of the fermion (boson) theory onto those of the boson (fermion) theory. Applications of the Fermi-Bose equivalence with the constructed Klein factors to well-known $(1+1)$ dimensional theories have been discussed. 
\end{abstract}




\maketitle

\section{Introduction}

One of the most powerful tools to analyze  $(1+1)$ dimensional interacting theories is the Fermi-Bose equivalence \cite{Mattis65, Mandelstam, coleman75} , 
which is also called as fermionization or bosonization in the literature \cite{emery79, solyom, Affleck88, fradkin, voit, stone94,delft, gogolin, senechal}.
In $(1+1)$ dimensions a fermion theory is mapped onto a boson theory completely by the fermionization procedure and a boson theory is mapped onto its counterpart fermion by the bosonization. The most useful advantage of the fermionization or the bosonization among others is that the strong coupling regimes of the theory can be mapped onto the weak coupling 
regimes of its counterpart theory. Often the counterpart theory turns out to be exactly solvable. 
For this reason the Fermi-Bose 
equivalence has been employed to study various subjects in both condensed matter physics and string theory. 
The applications of the Fermi-Bose equivalence range from various important subjects in condensed matter physics, 
such as the Hubbard model \cite{Hubbard57, Hubbard58, Hubbard63, essler}, the anisotropic Heisenberg spin-$1/2$ model \cite{Heisenberg, Jordan, Sirker}, the Kondo problem \cite{Hewson, Saleur1998, Saleur2000, Affleck:1990by}, the Tomonaga-Luttinger liquid \cite{Tomonaga, Luttinger, Haldane, fisher96, senechal, chang}, junctions of quantum wires \cite{chamon, oshikawa2005} and to those in string theory, which include the string partition functions on Riemann surfaces 
\cite{verlinde87, Eguchi, alvarez}, the string field theory \cite{Hlousek}, and the rolling tachyon \cite{Sen:2002nu, senreview, Lee:2005ge, Hassel, Tlee:06, TLee;08}. 

For the theory with a fermion (or boson) of a single species the equivalence is well established by the Mattis-Mandelstam formula \cite{Mattis65, Mandelstam}. By the formula fermion field operators are expressed in terms of the boson field operators in such a way that the excited states in the fermion theory can be represented as coherent states created by boson field operators. The Fermi-Bose equivalence has been the main tool for studying the non-perturbative properties of (1+1) dimensional theories since the seminal paper by Coleman \cite{coleman75}, where the fermion theory of the massive Thirring model is shown to be equivalent to the boson theory of the sine-Gordon model. The novel feature of the Mattis-Mandelstam formula is that the fermion field operators, represented by the boson field operators only, satisfy the Fermi statistics correctly. 
However, if the theory contains fermion fields of more than two species, one immediately finds that the Mattis-Mandelstam does not work if extended naively. Since the boson field operators of different species are 
independent of each other, 
they commute, hence the fermion field operators constructed by the corresponding boson field operators. The fermi statistics between different species is not reproduced by the fermi field operators constructed in terms of the boson field operators by the Mattis-Mandelstam formula. In order to remedy this drawback we may modify the Mattis-Madelstam formula by introducing additional operators in front of the fermion operators, which are termed as Klein factors. In order to ensure the anti-commutation relations between the fermion operators correctly, they have to satisfy certain conditions. In the literature one may find various useful discussions on the construction of the Klein factors and their applications. It may not be difficult to make the fermion operators of different species anticommute each other by introducing the Klein factors which have some simple structures. For an example, to represent them explicitly, one may choose the Dirac gamma matrices, which satisfy the Clifford algebra. 

But if the theory contains some non-trivial interactions, 
the bosonized or fermionized action still contains Klein factors. 
Although these Klein factors may not appear explicitly in evaluation of some physical quantities, 
their appearances in the resultant action could be sources of controversy in some cases. We may encounter a similar difficulty when we try to apply the boundary state formulation \cite{callan90, Polchinski:1994my} to the (1+1) dimensional theories in condensed matter physics and string theory. The boundary state formulation is one of the most efficient framework to calculate the correlation functions of physical operators
in the field theory defined on the one dimensional space with a boundary. In condensed matter physics many of the aforementioned $(1+1)$ dimensional models are defined on the one dimensional space with a boundary. 
In string theory, the open 
strings attached on D-branes are described by $(1+1)$ dimensional worldsheet action defined on the one dimensional space with boundaries. The simple boundary states such as the Neumann state and the Dirichlet state are of 
course well defined both in the boson theory and in the fermion theory. However, the boundary states of the boson 
theory and those of its corresponding fermion theory are not mapped precisely onto each other by 
the Fermi-Bose equivalence. The Klein factors enter.  
Suppose that we define those simple boundary states in a boson theory
and apply the fermionization with the conventional representations of the Klein factors to the boundary states.
Since the boundary conditions of the boson theory are linear in terms of the boson fields, 
the Klein factors explicitly enter the boundary conditions. 

In recent papers, in order to resolve those difficulties, associated with the conventional representations of the Klein factors, a new representation of the Klein factors has been proposed and applied to the critical boundary sine-Gordon model \cite{Hassel}, the rolling tachyon \cite{Lee:2005ge, Tlee:06, TLee;08}, and the quantum Brownian motion on a triangular lattice \cite{Lee2009q}. Since the new representation of Klein factors only utilize the zero modes of the boson field operators and reduce to $c$-numbers upon the simple boundary conditions, we can map the boundary states of the boson theory onto those of the fermion theory precisely and vice vera.  
In this paper we extend this construction to the most general cases and present an explicit expression of Klein 
factors. Then we apply them to some well-known $(1+1)$ dimensional field theories, which include $SU(2)$ Thirring model, the Gross-Neveu models with $N=2$ and the chiral Gross-Neveu model with $N=2$ and $N=3$. Some of the results are new.

\section{Klein Factors and Boundary States}

We begin with the Mattis-Madelstam formula for the fermion field of a single species. Let us consider a left moving free boson field in the $(1+1)$ dimensional Euclidean space-time, $\phi_L(\t,\s) = \phi_L(\tau+i\sigma)$. With a periodic boundary condition $\phi_L(\t,\s + 2\pi) = \phi_L(\t,\s)$, it 
may be written in terms of the oscillator operators as follows
\beq
 \phi_L(\tau+i\sigma)=
 \frac{1}{\sqrt{2}} x_L- \frac{i}{\sqrt{2}} p_L(\tau+i\sigma)+  \sum_{n=1}\frac{i}{\sqrt{2n}}
 \left(a_n e^{-n(\tau+i\sigma)} - a^\dag_ne^{n(\tau+i\sigma)} \right) , \label{leftboson}
\eeq
where the fundamental boson commutation relations are
\beq
 \left[ x_L,p_L\right]=i,~~~\left[ a_m, a^\dag_n \right] = \delta_{mn}.
\eeq  
The Mattis-Madelstam formula states that the free fermion field operator $\psi_L$ and $\psi^\dag_L$, satisfying 
anticommutation relations
\beq
\{\psi_L(\s), \psi_L(\s^\prime)\} = 0 , \quad 
\{\psi^{ \dagger}_L(\s), \psi_L(\s^\prime)\} =  2\pi \d(\s-\s^\prime), \label{antia}
\eeq
in the $(1+1)$ dimensions can be represented in terms of the boson field operators as follows
\beq
\psi_L = :e^{-i\sqrt{2} \phi_L}:, \quad \psi^\dag_L = :e^{i\sqrt{2} \phi_L}: . \label{leftfermion}
\eeq
Now let us introduce a right moving free boson field, $\phi_R(\t,\s)= \phi_R(\tau-i\sigma)$
\beq
\phi_R(\tau-i\sigma)&=&\frac{1}{\sqrt{2}} x_R- \frac{i}{\sqrt{2}} 
p_R(\tau-i\sigma)+\sum_{n=1}\frac{i}{\sqrt{2n}}
\left(\tilde a_n e^{-n(\tau-i\sigma)} - \tilde a^\dag_ne^{n(\tau-i\sigma)} \right)
\eeq
with the fundamental boson commutation relations
\beq
\left[ x_R,p_R\right]=i,~~~\left[ \tilde a_m, \tilde a^\dag_n \right] = \delta_{mn}. 
\eeq 
We may define the right moving fermion field in terms of the right moving boson as in the case of the left moving fermion
field 
\beq
\psi_R = :e^{-i\sqrt{2} \phi_R}:,\quad \psi^\dag_R = :e^{i\sqrt{2} \phi_R}: .\label{rightfermion}
\eeq
They satisfy the anticommutation relations of the fermion fields
\beq
\{\psi_R(\s), \psi_R(\s^\prime)\} = 0 , \quad 
\{\psi^{ \dagger}_R(\s), \psi_R(\s^\prime)\} =  2\pi \d(\s-\s^\prime). \label{antib}
\eeq

At this point we already encounter the difficulty associated with the bosonization: 
Since the left moving boson field operator and 
the right moving boson field operators are completely independent of each other, they commute
\beq
[\phi_L(\s), \phi_R(\s^\prime) ] = 0.
\eeq
It follows from this that the left moving fermion field operator and the right moving fermion field 
operator defined by Eqs.(\ref{leftfermion},\ref{rightfermion}) satisfy the commutation relations instead of the 
anticommutation relations. We need to introduce additional factors $\eta_L$, $\eta_R$ in front of 
the fermion operators 
\beq
\psi_L &=& \eta_L :e^{-i\sqrt{2} \phi_L}:, \quad 
\psi_R = \eta_R :e^{i\sqrt{2} \phi_R}:, \label{kleinfermion}
\eeq
so that the anti-commutation relations between fermion operators are reproduced
\beq
\{\psi_L(\s), \psi_R(\s^\prime)\} &=& 0 , \quad 
\{\psi^{ \dagger}_L(\s), \psi_R(\s^\prime)\} = 0, \quad 
\{\psi_L(\s), \psi^\dag_R(\s^\prime)\} = 0, \quad \{\psi^\dag_L(\s), \psi^\dag_R(\s^\prime)\} = 0 . \label{antic}
\eeq
Note that if the Klein factors $\eta_{L/R}$ commute with the boson operators $\phi_{L/R}$
and satisfy the Clifford algebra 
\beq
\{\eta_i, \eta_j\} = 2 \delta_{ij},  \quad i, j = L, R, 
\eeq
the constructed fermion operators Eq.(\ref{kleinfermion}) satisfy 
the correct anticommutation relations among the fermion operators Eqs.(\ref{antia}, \ref{antib}, \ref{antic}).
One may represent them by the Pauli matrices. To be explicit, we choose
\beq 
\eta_L = \s_1,\qquad  \eta_R = \s_2.
\eeq 
For the theory with fermion fields of more than two species, in general one may choose the Dirac matrices to represent the Klein factors. 
This is one of the conventional methods to construct the Klein factors. With this representation of the Klein factors,
the anti-commutation relations between the fermion operators are easily reproduced, 
but the Klein factors make their explicit appearances in the bosonized action. For an example, 
consider the Dirac mass term in the fermion theory $\bar \psi \psi$,
which may be written in terms of the boson field operators as 
\beq
\bar \psi \psi = \psi^\dag_L \psi_R + \psi^\dag_R \psi_L = \eta^\dag_L \eta_R :e^{\sqrt{2} i \phi}: 
+ \eta^\dag_R \eta_L :e^{-\sqrt{2}i \phi}: =  i\s_3 \left(:e^{\sqrt{2} i \phi}: - :e^{-\sqrt{2}i \phi}: \right) .
\eeq 
It is not clear how to deal with the Klein factors in the bosonized action. Often the Klein factors in the actions 
are ignored. Or one of the eigenstates of the Klein factors is chosen, while others are projected out. It is certainly 
ambiguous. For the example at hand, among 
\beq
\pm i\left(:e^{\sqrt{2} i \phi}: - :e^{-\sqrt{2}i \phi}: \right) 
\eeq
which boson operators we would choose to represent the Dirac mass term ?


A similar difficulty is encountered when we construct the boundary states. The Neumann boundary condition and the Dirichlet boundary condition are simply given in terms of boson field operators as 
\begin{subequations}
\beq
\phi_L|N\rangle &=& \phi_R|N\rangle, \label{neumann1}\\
\phi_L|D\rangle &= &-\phi_R|D\rangle . \label{dirichlet1}
\eeq
\end{subequations}
But if we apply the fermion representations Eq.(\ref{kleinfermion}) to express the boundary condition in terms of fermion operators, we realize that it cannot be written in terms of the fermion fields only
\begin{subequations}
\beq
\psi_L|N\rangle &=& \eta_L \eta_R^\dag \psi^\dagger_R|N\rangle = i \s_3 \psi^\dagger_R|N\rangle,
~~\psi^\dagger_L|N\rangle = \eta_L^\dag \eta_R \psi_R|N\rangle = i \s_3 \psi_R|N\rangle, \\
\psi_L|D\rangle &=& \eta_L \eta_R \psi_R|D\rangle = i \s_3 \psi_R|D\rangle,
~~\psi^\dagger_L|D\rangle = \eta^\dag_L \eta^\dag_R \psi_R^\dag|D\rangle = i \s_3 \psi_R^\dag|D\rangle .
\eeq
\end{subequations}
The Klein factors enter the boundary conditions explicitly. The free boson field theory defined on a space-time with a boundary is not mapped precisely onto a free fermion field theory by the fermi-bose equivalence with the Klein factors represented by the Pauli matrices.

Other conventional representations of  Klein factors also 
lead us to the same conclusion. An alternative representation \cite{senechal} of the Klein factors is
\beq
\eta_i = \exp\left(i \pi \sum_{i < j} p_j \right), ~~~ i,\, j = 1, 2, \dots, N. 
\eeq
For the single boson theory, we may write
\beq
\psi_L = :e^{-i\sqrt{2} \phi_L}:, ~~
\psi_R = e^{i \pi p_L} :e^{i\sqrt{2} \phi_R}:.
\eeq
The Klein factors in this representation do not commute with the Mattis-Mandelstam operators, but commute with
each other, in contrast to the Dirac matrix representation of the Klein factors. 
In fact, the anti-commutation relations between the fermion 
field operators are ensured by the non-commutative algebra between the Klein factors and the Mattis-Madelstam 
operators,
\beq
\{ \psi_L(\s), \psi_R(\sp) \} 
=\{ e^{-i\sqrt{2} \phi_L}, e^{i\pi p_L} \} :e^{i\sqrt{2} \phi_R}: =0.
\eeq
Here we make use of the Baker-Campbell-Hausdor formula
\beq
e^A e^B = e^B e^A e^{[A,B]}, ~~~~\text{if}~~ [A,B] = c\text{-number}.
\eeq
Although the anti-commutation relations between the fermion field operators are satisfied, the difficulties 
associated with the Dirac matrix representation of the Klein factors still remain unresolved. If we bosonize the 
Dirac fermion mass term by using this representation, we have 
\beq
\bar \psi \psi = \psi^\dag_L \psi_R + \psi^\dag_R \psi_L 
= - e^{i\pi p_L} :e^{i\sqrt{2} \phi}: + e^{-i\pi p_L} :e^{-i\sqrt{2} \phi}:.
\eeq 
The Klein factors $e^{\pm i \pi p_L}$ enter explicitly. The same problem occurs when we map the 
simple boundary states. 
The Neumann boundary condition in the boson theory Eq.(\ref{neumann1})
is now expressed in term of the fermion field operators
as
\beq
\psi_L |N \rangle = e^{i\pi p_L} \psi^\dagger_R |N\rangle, ~~
\psi^\dagger_L |N\rangle = e^{-i\pi p_L} \psi_R |N\rangle . 
\eeq
Note that the Klein factors enter the boundary condition as before. Thus, this alternative 
representation of the Klein factors also suffers the same problems.  

In recent works, a new representation of the Klein factors has been proposed. The new representation only 
utilizes the zero modes of the momentum operators $p_L$ and $p_R$. 
In the case of the theory with a single boson the most general 
forms of the Klein factors may be written as 
\beq  \psi_{L} = e^{- \frac{\pi i}{2}(\a^L p_L + \b^L p_R)}
e^{-\sqrt{2}i X_L}, ~~\psi_{R} &=& e^{\frac{\pi i}{2} (\a^R p_L + \b^R p_R)}
e^{\sqrt{2}i X_R}
\eeq  
where $\a^{L}, \a^{R}, \b^{L}, \b^{R}$ are constant parameters, to be fixed by suitable 
conditions. 
We may derive the conditions for the Klein factors to satisfy, requiring followings: 
\begin{itemize}
\item
The fundamental anti-commutation relations between fermion operators are properly reproduced.
\item 
The boundary conditions for the simple boundary states of the boson theory are
mapped onto those of the corresponding fermion theory by the Fermi-Bose equivalence.
\item
Relevant interaction terms of the fermion theory, including the Dirac mass term, should be mapped onto the 
corresponding terms in the boson theory, which do not contain any Klein factor.
\end{itemize}
It is not difficult to find a solution satisfying all those conditions for the theory with a single boson: 
\beq   
\psi_{L} = e^{-\frac{\pi i}{2}(p_L+ p_R)} e^{-\sqrt{2}i \phi_L}, \quad
\psi_{R} = e^{-\frac{\pi i}{2} (p_L+p_R)} e^{\sqrt{2}i \phi_R}. 
\eeq
One can easily check that the anti-commutation relations Eqs.(\ref{antia}, \ref{antib}, \ref{antic})
between the fermion field operators are satisfied. The Dirac mass operator in the fermion theory is now 
unambiguously mapped onto the periodic potential of the boson theory
\beq
\bar \psi \psi = :e^{i\sqrt{2} \phi}: +  :e^{-i\sqrt{2} \phi}: = 2: \cos (\sqrt{2} \phi): .
\eeq
The Neumann boundary condition of the boson theory Eq.(\ref{neumann1}) is mapped onto the following linear 
condition in terms of the fermion field operators 
\beq   
\psi_L|N\rangle = i \psi^{\dagger}_R |N\rangle,\quad
\psi_L^\dagger|N\rangle = i \psi_R|N\rangle.
\eeq
Thus, at least for the theory of a single boson field, we have a proper representation of the Klein factors,
which is free from the associated problems. 
The Klein factors for the case of the model with two boson fields have been constructed \cite{Lee:2005ge} 
and applied to the rolling tachyon \cite{Hassel, TLee;08}
in string theory . The new representation of the Klein factors also has been applied to the quantum Brownian 
motion on a triangular lattice, which requires three boson fields \cite{Lee2009q}.

We may recall 
the kink or soliton operator in the anisotropic Heisenberg $(XXZ)$ model \cite{fradkin}, which changes 
the statistics of the spin-1/2 operator. By the Jordan-Wigner transformation \cite{Jordan} the spin-1/2 operators 
can be mapped to spinless fermions. Bu the fermion operators on different sites anticommute while the spin-1/2 operators on different sites commute.
In order to give the correct statistics to the fermion operators, one may introduce the kink operator,
which is formed of the total fermion number operator. The role of the kink operator is very much similar to 
that of the Klein factors. The zero modes of the boson momentum operators are in fact the total number operators in fermion theory
\beq
p_L =  \frac{1}{2\pi} \int^{2\pi}_0 \psi^\dag_L \psi_L d\s, ~~~ 
p_R =  -\frac{1}{2\pi} \int^{2\pi}_0 \psi^\dag_R \psi_R d\s ,
\eeq
which do not commute with the boson fields. The new representation of the Klein factors, which we are about to 
construct, may be considered as 
generalized kink operators, which have a more complex structure. 

\section{Conditions for Klein Factors}

We may write the most general form of Klein factors for the fermion fields 
which correspond to the $N$ boson fields as
\begin{subequations}
\beq
\psi^a_L &=& e^{-\frac{\pi}{2} i \sum_b\left(
\a^L_{ab} p^b_L + \b^L_{ab} p^b_R \right)} e^{-\sqrt{2} i \phi^a_L},~~
\psi^a_R = e^{\frac{\pi}{2} i \sum_b\left(
\a^R_{ab} p^b_L + \b^R_{ab} p^b_R \right)} e^{\sqrt{2} i \phi^a_R} \\
\psi^{a\dag}_L &=& e^{\sqrt{2} i \phi^a_L} e^{\frac{\pi}{2} i \sum_b\left(
\a^L_{ab} p^b_L + \b^L_{ab} p^b_R \right)},~~
\psi^{a\dag}_R = e^{-\sqrt{2} i \phi^a_R}e^{-\frac{\pi}{2} i \sum_b\left(
\a^R_{ab} p^b_L + \b^R_{ab} p^b_R \right)} 
\eeq
\end{subequations}
where $a, b = 1, 2, \dots, N$ and $\a^L_{ab}, \a^R_{ab}, \b^L_{ab}, \b^R_{ab}$ are constant parameters to be fixed by
the conditions, imposed on the Klein factors. 
The left moving boson field operators and the right moving ones may 
be expanded in terms of the oscillator modes as follows:
 \begin{subequations}
 \label{expan}
 \begin{eqnarray}
 \phi^a_L(\tau+i\sigma)&=&
 \frac{1}{\sqrt{2}} x^a_L-\frac{i}{\sqrt{2}} p^a_L(\tau+i\sigma)+\sum_{n=1
 }\frac{i}{\sqrt{2n}} \left(a^a_n e^{-n(\tau+i\sigma)}- a^{a\dag}_n e^{n(\tau+i\sigma)}\right),\label{expan:a} \\
 \phi^a_R(\tau-i\sigma)&=&\frac{1}{\sqrt{2}} x^a_R- \frac{i}{\sqrt{2}} 
 p^a_R(\tau-i\sigma)+\sum_{n=1}\frac{i}{\sqrt{2n}}\left(\tilde a^a_n e^{-n(\tau-i\sigma)}- \tilde a^{a\dag}_n
 e^{n(\tau-i\sigma)} \right), \label{expan:b}
 \end{eqnarray}
 \end{subequations}
with the non-vanishing fundamental commutators
\begin{subequations}
\beq
[x^a_L, x^b_L]&=& i \d^{ab}, ~~~[x^a_R, x^b_R] = i \d^{ab},  \\
\left[ a^a_m, a^{b\dag}_n \right] &=& \d^{ab}\d_{mn}, ~~~
\left[\tilde a^a_m, \tilde a^{a\dag}_n \right] = \d^{ab}\d_{mn}.
\eeq
\end{subequations}

\subsection{Anti-Commutation Relations between the Fermion Operators}

The first conditions to be imposed on the Klein factors are obtained by requiring the anti-commutation relations 
between fermion operators
\begin{subequations}
\beq
\{\psi^a_L(\s), \psi^b_L(\s^\prime)\} &=& 0 , \quad 
\{\psi^a_L(\s), \psi^b_R(\s^\prime)\} = 0, \label{anticomm1}\\
\{\psi^a_R(\s), \psi^b_R(\s^\prime)\} &=& 0, \quad
\{\psi^{a \dagger}_L(\s), \psi^b_R(\s^\prime)\} =0, \label{anticomm2}\\
\{\psi^{a \dagger}_L(\s), \psi^b_L(\s^\prime)\} &=& \d^{ab} 2\pi \d(\s-\s^\prime), \label{anticomm3}\\
\{\psi^{a\dagger}_R(\s), \psi^b_R(\s^\prime)\} &=& \d^{ab} 2\pi \d(\s-\s^\prime). \label{anticomm4}
\eeq
\end{subequations}
By some algebra we find that the anti-commutation relations between the fermion field operators
$\psi_{aL},~\psi_{aL},~\psi^{a \dagger}_L,~ \psi^{a \dagger}_R$, are ensured
if the following conditions are satisfied
\begin{subequations}
\beq 
e^{\frac{\pi i}{2}(\a^L_{ab}-\a^L_{ba})} &=& -1,~~{\rm for}~ a\not=b,\label{anti1}\\
e^{\frac{\pi i}{2}(\b^R_{ab}-\b^R_{ba})} &=& -1,~~{\rm for}~ a\not=b, \label{anti2}\\
e^{\frac{\pi i}{2}(\a^R_{ab}-\b^L_{ba})} &=& -1.\label{anti3}
\eeq
\end{subequations}
We note that the anti-commutation relations between the fermion operators alone 
cannot fix the Klein factors. We only have $(2N^2 - N)$ conditions to fix $4N^2$ parameters. 
Additional conditions would be obtained by requiring that
the simple boundary states are linearly represented in terms of the fermion
fields and the interaction terms are also uniquely represented in terms of the fermion fields
only in the fermion theory.

\subsection{The Simple Boundary States and Klein Factors}

Often the condensed matter systems and the string theory are defined on one dimensional space with a boundary. 
The boundary conditions for simple boundary states such as 
$\vert {\bf N} \rangle =|N,\dots,N\rangle$ and 
$\vert {\bf D}\rangle =|D,\dots,D\rangle$ should be realized linearly in terms of the fermion operators 
without the Klein factors. In condensed matter system interactions between physical fields defined on 
the bulk space and the impurities on the boundary are the main subjects of study. In string theory the 
open strings attached on D-branes are described by the $(1+1)$ dimensional field theory defined on one dimensional space with a boundary. Thus, when we apply the bosonization or the fermionization to these theories we encounter 
the difficulty, similar to those discussed in the previous section if the Klein factors enter the boundary 
condition explicitly. This requirement will impose some conditions for the Klein factors and yield the 
relations between $\a^L_{ab}, \a^R_{ab}, \b^L_{ab}, \b^R_{ab}$. 
We begin with the boundary state $\vert {\bf N} \rangle$.  

\subsubsection{The Neumann Boundary State} 

If the theory does not have any non-trivial interaction both in the bulk and the boundary, the fields
should be subject to the Neumann condition. 
The boundary condition for the state $\vert {\bf N} \rangle$  is given linearly in terms of 
the bosonic operator as
\beq   
\phi^a_L\vert {\bf N} \rangle = \phi^a_R\vert {\bf N} \rangle,\quad a = 1, 2, \dots, N,  
\eeq
which can be read in terms of normal modes as
\beq  x^a_L \vert {\bf N} \rangle = x^a_R\vert {\bf N} \rangle, \quad 
p^a_L\vert {\bf N} \rangle= - p^a_R\vert {\bf N} \rangle,\quad 
a^a_n\vert {\bf N} \rangle= - \tilde a^{a\dag}_{n}\vert {\bf N} \rangle .\nn  
\eeq
Applying the fermion field operator $\psi^a_L$ on the Neumann boundary state
\beq
\psi^a_L\vert {\bf N} \rangle 
&=& e^{\frac{\pi}{2}i \b^L_{aa}} \psi^\dagger_{aR}
e^{\frac{\pi}{2}i \sum_b \left(\a^R_{ab} - \b^R_{ab} -\a^L_{ab} + \b^L_{ab} \right)p^b_L} 
\vert {\bf N} \rangle ,
\eeq
we find that the Neumann boundary condition can be expressed linearly in 
terms of the fermion operators if the following conditions are satisfied 
\beq \label{nnn}
\a^L_{ab} -\a^R_{ab}  - \b^L_{ab} + \b^R_{ab} = 0.
\eeq

Under the condition Eq.(\ref{nnn}) we write the Neumann boundary condition in the fermion
theory as
\beq
\psi^a_L \vert {\bf N} \rangle = e^{\frac{\pi}{2}i \b^L_{aa}}\psi^\dagger_{aR}
\vert {\bf N} \rangle, \quad
\psi^{a \dagger}_L\vert {\bf N} \rangle
= e^{-\frac{\pi}{2}i(\a^L_{aa}-\b^L_{aa}+\b^R_{aa})}\psi^a_R
\vert {\bf N} \rangle. \label{neumann}
\eeq
These fermion boundary conditions should be also consistent with the fundamental anti-commutation relations between the fermion field operators Eqs.(\ref{anticomm1}, \ref{anticomm2}, \ref{anticomm3}, \ref{anticomm4}), 
\beq
\{\psi^{a\dag}_L(\s), \psi^{b}_L(\s^\prime)\}\vert {\bf N} \rangle 
&=& - e^{\frac{\pi}{2}i \b^L_{bb}} e^{-\frac{\pi}{2}i(\a^L_{aa}-\b^L_{aa}+\b^R_{aa})} 
\{\psi^{b \dagger}_R(\s^\prime), \psi^a_R(\s)\}\vert {\bf N} \rangle \nn\\
&=&\d^{ab} 2\pi \d(\s-\s^\prime)\vert {\bf N} \rangle .
\eeq
If $a\not=b$, LHS=RHS=0. So it does not requires any additional condition. But if $a=b$, 
the phase factor on RHS must be 1. It yields the following additional conditions
\beq \label{neumann2}
e^{-\frac{\pi}{2}i(\a^L_{aa}-2\b^L_{aa}+\b^R_{aa})} = -1, \quad a=1, 2, \dots, N.
\eeq

\subsubsection{The Dirichlet Boundary State}

In string theory the end points of the string coordinate fields in the directions, orthogonal to the worldsheet 
of the D-branes satisfy the Dirichlet condition. In the condensed matter systems, if the interactions 
between the impurities on the boundary and the physical fields on the bulk space become strong, 
the fixed point of the RG (renormalization group) flow is the Dirichlet state. 
We require that the Dirichlet boundary condition in the boson theory must be also mapped onto 
that in the fermion theory by the fermionization as in the case of the Neumann boundary condition.
The boundary condition for $\vert {\bf D}\rangle$ is given in terms of the bosonic operator as 
\beq
\phi^a_L \vert {\bf D}\rangle = - \phi^a_R \vert {\bf D}\rangle, ~~~ a = 1, 2, \dots, N.
\eeq
If it is written in terms of normal modes,
\beq  x^a_L\vert {\bf D}\rangle= -x^a_R\vert {\bf D}\rangle, \quad 
p^a_L\vert {\bf D}\rangle=  p^a_R\vert {\bf D}\rangle,\quad 
a^a_n\vert {\bf D}\rangle=  \tilde a^{a\dag}_{n}\vert {\bf D}\rangle. \nn  
\eeq
Since it may be written in terms of fermion operator as
\beq
\psi^a_L\vert {\bf D}\rangle
&=& e^{-\frac{\pi}{2}i\left(\b^L_{aa}+ \b^R_{aa}\right)} \psi^a_R e^{-\frac{\pi}{2}i \sum_b\left(\a^L_{ab}+
\a^R_{ab} + \b^L_{ab} + \b^R_{ab}\right)p^b_L}\vert {\bf D}\rangle, 
\eeq
we should impose the following condition
\beq \label{ddd}
\a^L_{ab}+ \a^R_{ab} + \b^L_{ab} + \b^R_{ab} = 0.
\eeq
Then the Dirichlet boundary condition is linearly represented
by the fermion field operators without the Klein factors
\beq \label{ddd1}
\psi^a_L\vert {\bf D}\rangle = e^{-\frac{\pi}{2}i\left(\b^L_{aa}+ \b^R_{aa}\right)} 
\psi^a_R\vert {\bf D}\rangle.
\eeq
Under the condition Eq.(\ref{ddd}), we also find 
\beq \label{ddd2}
\psi^{a\dagger}_L\vert {\bf D}\rangle 
= e^{-\frac{\pi}{2}i \left(\a^L_{aa}+ \b^L_{aa}\right)} \psi^{a\dagger}_R \vert {\bf D}\rangle.
\eeq
These two boundary conditions Eq.(\ref{ddd1}) and Eq.(\ref{ddd2})
should be compatible with the fundamental fermion anti-commutation relations. It follows 
from this requirement that
\beq \label{cons2}
e^{-\frac{\pi}{2}i \left(\a^L_{aa} + 2\b^L_{aa} + \b^R_{aa}\right)} = -1,\quad a = 1, 2, \dots, N.
\eeq

{\bf Other boundary states and Klein factors}:
We may repeat the same procedure for other mixed boundary states such as 
$|D,N, \dots, N\rangle$ or $|N,\dots,D\rangle$. However, it does not produce any additional condition for the Klein factors, since they are all related by $SU(N)$ global transformations. 

\subsection{The Periodic Potential and  Fermion Mass Term}

The boson theory in $(1+1)$ dimensions may contain the periodic potential terms, 
\beq
U_a + U^\dag_a = :e^{i\sqrt{2} \phi^a}: + :e^{-i\sqrt{2} \phi^a}:, ~~~ a =1, 2, \dots, N ,
\eeq
which arise in the bulk or on the boundary. In a condensed matter theory these terms correspond to the 
bosonized form of the Umklapp process \cite{fradkin}. 
When fermionizing the periodic potential terms, we may rewrite them as bilinear terms in 
fermion operators
\beq
:e^{i\sqrt{2} \phi^a}: &=& :e^{i\sqrt{2}\left(\phi^a_L+ \phi^a_R\right)}: \nn\\
&=& \psi^{a\dag}_L e^{-\frac{\pi}{2} i \sum_b\left(\a^L_{ab} p^b_L + \b^L_{ab} p^b_R \right)}
e^{-\frac{\pi}{2} i \sum_b\left(\a^R_{ab} p^b_L + \b^R_{ab} p^b_R \right)}\psi^a_R.
\eeq
We do not want the Klein factors to enter explicitly in the fermion bilinear operators. It follows 
from this constraint that
\beq \label{mass}
\a^L_{ab} + \a^R_{ab} =0, \quad \b^L_{ab} + \b^R_{ab} =0, ~~~ a, b = 1, 2, \dots, N .
\eeq
Under these conditions they could be interpreted as the Dirac mass terms
\beq
U_a + U^\dag_a = \psi^{a\dag}_L \psi^a_R + \psi^{a\dag}_R \psi^a_L = \bar \psi^a \psi^a ,
\eeq
of the relativistic Dirac fermion theory. 

\subsection{Interactions and The Kein Factors}

When we apply the fermionization or the bosonization to $(1+1)$ dimensional theories, we often find that 
non-trivial Klein factors also appear in the interaction terms of the counterpart theories. These Klein factors 
are sources of controversy. Since there are still some rooms to impose additional conditions for the Klein factors, 
we may impose further conditions consistently to get rid of those Klein factors. Let us consider following interaction terms between boson fields
\beq
V_a = V_{a,a+1}+ V_{a,a+1}^\dagger= 
:e^{\frac{i}{\sqrt{2}}(\phi^a-\phi^{a+1})}:+ :e^{-\frac{i}{\sqrt{2}}(\phi^a-\phi^{a+1})}:, 
\eeq
where $a=1, 2, \dots, N, \quad \phi^{N+1} = \phi^1$. The interaction terms of this type arise in the theory of junctions of quantum wires as boundary interactions. In $(1+1)$ dimensional quantum field theory  interaction terms of this type also appear in the bosonized bulk action of the chiral Gross-Neveu model. If these terms arise in the bulk,
we may scale the boson fields as $\phi^a \rightarrow 2 \phi^a$ 
\beq
V_a = V_{a,a+1}+ V_{a,a+1}^\dagger= 
:e^{i\sqrt{2}(\phi^a-\phi^{a+1})}:+ :e^{-i\sqrt{2}(\phi^a-\phi^{a+1})}:. 
\eeq
Then by using the condition Eq.(\ref{mass}), we may rewrite them as four fermi terms in the fermion theory without
nontrivial Klein factors
\beq
V_a = U^{a+1 \dag} U^a + U^{a \dag} U^{a+1} = \left(\psi^{a+1\dag}_R \psi^{a+1}_L \right)
\left(\psi^{a\dag}_L \psi^{a}_R \right) + \left(\psi^{a\dag}_R \psi^a_L \right)\left(\psi^{a+1 \dag}_L \psi^{a+1}_R \right) . 
\eeq
We do not need to impose additional conditions to fermionize the bulk interaction terms of this type.

But if the interaction terms of this type arise on the boundary, we should treat them with some care. On the boundary they cannot be written as four fermi terms. If we turn off the interaction, the boundary state should reduce to the Neumann state. And the boundary state $|{\bf B} \rangle$ in the presence of the interaction is
obtained by applying the boundary interaction terms on the simple boundary state $\vert {\bf N}\rangle$
\beq
\vert {\bf B} \rangle &=& V_a\vert {\bf N} \rangle \nn\\
 &=& \left( U^{a+1 \dag} U^a + U^{a \dag} U^{a+1} \right)\vert {\bf N} \rangle \nn\\
 &=& \left[\left(\psi^{a+1\dag}_R \psi^{a+1}_L \right)
\left(\psi^{a\dag}_L \psi^{a}_R \right) + \left(\psi^{a\dag}_R \psi^a_L \right)
\left(\psi^{a+1 \dag}_L \psi^{a+1}_R\right) \right] \vert {\bf N} \rangle .
\eeq
If we make use of the Neumann condition Eq.(\ref{neumann}), we find that $U^a \vert {\bf N} \rangle$ is vanishing
\beq
U^a \vert {\bf N} \rangle &=& \left(\psi^{a\dag}_L \psi^{a}_R \right) \vert {\bf N} \rangle \nn\\
&=& e^{\frac{\pi}{2}i \left(\a^L_{aa}-\b^L_{aa}+\b^R_{aa}\right)} \left( \psi^{a\dag}_L \psi^{a\dag}_L \right)
\vert {\bf N} \rangle \nn\\
&=& 0 ,
\eeq 
thanks to the fermi statistics. Thus, if we rewrite the interaction term $V_a$ as a four fermi term on the 
boundary, it becomes a null operator. 

On the boundary the interaction term $V_a$ may be represented as a bilinear operator of fermi fields instead. 
The boundary interaction term
$V_{a,a+1}$ acting on the Neumann boundary state
may be written in terms of the left moving fermion field operators as
\beq
V_{a,a+1} \vert {\bf N} \rangle &=&  e^{-\frac{\pi}{2}i\left(\a^L_{aa} - \a^L_{a+1,a}\right)} \psi^\dagger_{a+1,L} \psi_{aL} e^{-\frac{\pi}{2}i \sum_b \left(\a^L_{a+1,b} - \a^L_{ab} -\b^L_{a+1,b} + \b^L_{ab}\right)p^b_L}
\vert {\bf N} \rangle.
\eeq
It follows that in order to remove non-trivial Klein factors in the fermion form of the interaction term 
we should impose the following condition
\beq \label{nnn1l}
\a^L_{a+1,b} - \a^L_{ab} -\b^L_{a+1,b} + \b^L_{ab} = 0, ~~a, b = 1, 2,\dots, N.
\eeq
Once this condition is imposed, the
boundary interaction term $V_a$ may be written as 
\beq
V_{a,a+1} \vert {\bf N} \rangle = e^{-\frac{\pi}{2}i\left(\a^L_{aa} - \a^L_{a+1,a}\right)} \psi^\dagger_{a+1,L} \psi_{aL}\vert {\bf N} \rangle.
\eeq
We note that the boundary term $V_{a,a+1}$ can be also equally written in terms of the 
right moving chiral fermion field operators as
\beq
V_{a,a+1} \vert {\bf N} \rangle
&=&  e^{-\frac{\pi}{2} i(\b^R_{a+1,a+1} - \b^R_{a,a+1})} \psi^{\dagger}_{aR}
\psi_{a+1,R} e^{\frac{\pi}{2}i \sum_b \left(\a^R_{ab} - \b^R_{ab} -\a^R_{a+1,b} + \b^R_{a+1b} \right)p^b_L} 
|{\bf N}\rangle. \nn
\eeq
Thus, if the following condition is satisfied
\beq \label{nnn1r}
\a^R_{ab} - \a^R_{a+1,b} - \b^R_{ab} + \b^R_{a+1,b} = 0,
\eeq
the boundary term can be written as a bilinear of the right moving fermion field operators
\beq
V_{a,a+1} \vert {\bf N} \rangle = e^{-\frac{\pi}{2} i(\b^R_{a+1,a+1} - \b^R_{a,a+1})} \psi^{\dagger}_{aR} \psi_{a+1,R} |{\bf N}\rangle .
\eeq

As we repeat the same procedure for the boundary interaction term $V_{a,a+1}^\dagger$, we obtain the fermion form of the boundary interaction term as
\beq
V_{a,a+1}^\dagger |{\bf N}\rangle = e^{-\frac{\pi}{2}i 
(\a^L_{a+1,a+1} - \a^L_{a,a+1})} \psi^\dagger_{aL} \psi_{a+1,L} |{\bf N}\rangle.
\eeq
A non-trivial Klein factor does not arise if the condition Eq.(\ref{nnn1l}) is satisfied.
Rewriting the boundary interaction term $V_a^\dagger$ in terms of the right moving fermion operators, we have
\beq
V_a^\dagger |{\bf N}\rangle  = e^{\frac{\pi}{2}i \left(\b^R_{a+1,a}-\b^R_{a,a}\right)}\psi^\dagger_{a+1,R} 
\psi_{a,R}|{\bf N}\rangle 
\eeq
under the condition Eq.(\ref{nnn1r}).
Therefore, we may write the boundary interaction term 
$V_a = V_{a,a+1}+V_{a,a+1}^\dagger$ in terms of the left moving fermion field operators as 
\beq
V_a |{\bf N}\rangle &=& 
\left(e^{-\frac{\pi}{2}i(\a^L_{a,a} - \a^L_{a+1,a})}\psi^\dagger_{a+1,L} \psi_{aL}
+ e^{-\frac{\pi}{2}i(\a^L_{a+1,a+1} - \a^L_{a,a+1})} \psi^\dagger_{a,L} \psi_{a+1,L}\right) |{\bf N}\rangle,
\eeq
or in terms of the right moving fermion field operators as
\beq
V_a |{\bf N}\rangle =
\left(e^{-\frac{\pi}{2}i(\b^R_{a+1,a+1} - \b^R_{a,a+1})}\psi^\dagger_{aR} \psi_{a+1,R}
+ e^{\frac{\pi}{2}i(\b^R_{a+1,a} - \b^R_{a,a})} \psi^\dagger_{a+1,R} \psi_{a,R}\right) |{\bf N}\rangle . 
\eeq
Note that, however, the fermion form of $V_a$ is not manifestly Hermitian unless appropriate conditions for
the Klein factors are imposed. It can be achieved by introducing the following conditions
\begin{subequations}
\beq 
e^{\frac{\pi}{2}i(\b^R_{a+1,a+1} - \b^R_{a+1,a} - \b^R_{a,a+1} + \b^R_{a,a})} &=& 1, \label{herm1}\\
e^{\frac{\pi}{2}i(\a^L_{a+1,a+1} - \a^L_{a+1,a} - \a^L_{a,a+1} + \a^L_{a,a})} &=& 1. \label{herm2}
\eeq
\end{subequations}

\section{Solutions}

In the previous section we exhaust the necessary conditions for the Klein factors to satisfy. If there exists a set of Klein factors, which satisfy all those conditions consistently, we can map the boson theories onto the corresponding fermion theories which do not contain any non-trivial Klein factors in their actions and vice versa. 
In this section we show that those conditions are consistent and there are many sets of solutions parameterized by 
some integers. At the end we present an explicit solution, which may be the simplest one. 

Making use of Eqs.(\ref{nnn},\ref{ddd}), we may replace all $\b^L_{ab}$ and $\b^R_{ab}$ by $\a^L_{ab}$ and $\a^R_{ab}$ in the conditions
\beq \label{ab}
\b^L_{ab} = -\a^R_{ab}, ~~~ \b^R_{ab} = -\a^L_{ab}, \quad a, \,b = 1, 2, \dots, N.
\eeq
Then the equations Eq.(\ref{nnn1l}) and Eq.(\ref{nnn1r}) reduce to the following conditions
\beq
\a^L_{a+1,b} -\a^L_{a,b}+\a^R_{a+1,b}-\a^R_{a,b} =0, \quad a,\, b = 1, 2, \dots, N.
\eeq
These conditions become redundant if the conditions Eq.(\ref{mass}) are chosen.
Under the conditions Eq.(\ref{ab}),  Eqs.(\ref{neumann2},\ref{cons2}) may be written as
$e^{-\pi i \a^R_{aa}}=-1$, {\it i.e.},
\beq
\a^R_{aa}= 2n^R_{aa} +1, \quad n^R_{aa} \in {\bf Z}, \quad a= 1, 2, \dots, N.
\eeq
The conditions Eqs.(\ref{anti1}, \ref{anti2}, \ref{anti3}) which ensure the anticommutation 
relations between fermion operators, $\psi^a_L$, $\psi^a_R$ reduce to 
\beq 
e^{\frac{\pi}{2}i(\a^L_{ab} -\a^L_{ba})} = -1, ~~
e^{\frac{\pi}{2}i(\a^R_{ab} + \a^R_{ba})} = -1  .
\eeq
These conditions are satisfied by rewriting the antisymmetric part of $(\a^L)$ and the symmetric part of 
$(\a^R)$ by integers $m^L_{ab}$ and $n^R_{ab}$
\beq \label{cond2}
\a^L_{ab} -\a^L_{ba} = 2(2m^L_{ab}+1), ~~~
\a^R_{ab} + \a^R_{ba} = 2(2n^R_{ab}+1), ~~a < b, ~~m^L_{ab}, \, n^R_{ab} \in {\bf Z}.
\eeq
With the conditions Eq.(\ref{mass}), Eq.(\ref{cond2}) is rewritten in terms of $\a^L_{ab}$ only
\beq \label{cond3}
\a^L_{ab} -\a^L_{ba} = 2(2m^L_{ab}+1), ~~~
\a^L_{ab} + \a^L_{ba} = -2(2n^R_{ab}+1).
\eeq
If a set of solution for the compoents of $(\a^L)$ is found, the components of $(\a^R)$,
$(\b^L)$, $(\b^R)$ are determined by
\beq
\a^L_{ab} = -\a^R_{ab} = \b^L_{ab} = -\b^R_{ab}.
\eeq
The diagonal components may be solved by some integers $n_a$
\beq
\a^L_{aa} =-\a^R_{aa} = \b^L_{aa} = -\b^R_{aa}= 2n_a+1, \quad n_a \in {\bf Z},
\eeq
where
$n_a = -n^R_{aa} -1$.
The Hermitianity conditions Eq.(\ref{herm1}, \ref{herm2}) set a condition to be satisfied by the 
integers $n_a$
\beq
e^{\pi i(n_{a+1}+ n_a)} = 1.
\eeq
Let us write the upper triangle components of $(\a^L)$ as 
\beq 
\a^L_{ab}= 2n_{ab} = 
2(m^L_{ab} - n^R_{ab}), ~~~\text{for}~~~ a<b
\eeq 
where $n_{ab}$ is an integer. Then the lower triangle components of $(\a^L)$ is written by 
\beq
\a^L_{ba} = \a^L_{ab} -4m^L_{ab} -2, ~~~\text{for}~~~ a<b .
\eeq
In summary, we have solutions parameterized by some integers $ n_{ab}, \, m_{ab},\, n_a $, $a<b$, 
\begin{subequations} 
\beq
\a^L_{ab} &=& 2 n_{ab}, ~~~ \a^L_{ba} = 2n_{ab} -2(m_{ab}+1),~~~\text{for}~~~ a<b , \label{general1}\\
\a^L_{aa} &=& 2n_a+1, ~~~e^{\pi i (n_{a+1} + n_a)}=1 ,  \label{general2}\\
\left(\a^L\right) &=& -\left(\a^R\right) = \left(\b^L\right) = -\left(\b^R\right). \label{general3}
\eeq
\end{subequations}

{\bf Some Simple Solutions}: The simplest choice may be $n_{ab} = 0$ for $a<b$ and $n_a=0$. The upper triangle of $(\a^L)$ is null and the diagonal components are 1. By choosing $m_{ab}=-1$, we can make the 
off-diagonal components of $(\a^L)$ have the same value of 2,
\beq
(\a^L) = 
 \begin{pmatrix}
  1 & 0 & \cdots & 0 \\
  2 & 1 & \cdots & 0 \\
  \vdots  & \vdots  & \ddots & \vdots  \\
  2 & 2 & \cdots & 1 
 \end{pmatrix} = -\left(\a^R\right) = \left(\b^L\right) = -\left(\b^R\right). \label{simple1}
\eeq
We can choose alternatively, $\a^L_{ab} =0$, $m^L_{ab} = -1$ for $a>b $ and $n_a =0$. This yields a solution for 
$(\a^L)$, of which lower triangle is null
\beq
(\a^L) = 
 \begin{pmatrix}
  1 & 2 & \cdots & 2 \\
  0 & 1 & \cdots & 2 \\
  \vdots  & \vdots  & \ddots & \vdots  \\
  0 & 0 & \cdots & 1 
 \end{pmatrix}= -\left(\a^R\right) = \left(\b^L\right) = -\left(\b^R\right).
\eeq
In the previous work \cite{Lee2009q} on the Brownian motion on a triangular lattice, we apply the $N=3$ solution,
of which explicit expression is given by
\beq
(\a^L) =
 \begin{pmatrix}
  1 & 2 & 0 \\
  0 & 1 & 2 \\
  2 & 0 & 1 
 \end{pmatrix}= -\left(\a^R\right) = \left(\b^L\right) = -\left(\b^R\right).
\eeq
It corresponds to the solution Eqs.(\ref{general1}, \ref{general2}) with
\beq
n_1 = n_2 = n_3 =0,~~~ n_{12}=1, ~~~ n_{13} =0, ~~~ n_{23}= 1, ~~
m_{12} =0,~~~ m_{23}=0, ~~~ m_{13} = -1.
\eeq

\section{Applications of Bose-Fermi Equivalence and Klein Factors}

Here we apply the set of Klein factors, which constructed in the last section, to some well-known models in 
quantum field theory and condensed matter physics. We begin with the model with a single boson field. 
As we discussed in the introduction we have some difficulties with the conventional representations of the Klein
factor even for the case of theories with a single boson field. Since in $(1+1)$ dimensions the left moving boson and the right moving boson are completely independent of each other, they must be treated as two independent species. 

\subsection{Theory with a Single Boson Field: N=1 System}

If the theory has only a single boson, the matrices $(\a^{L/R})$ and $(\b^{L/R})$ reduce to numbers. The 
simple solution Eq.(\ref{simple1}) is written as 
\beq
\a^L = 1,~~ \a^R = -1,~~ \b^L =1,~~ \b^R = -1.
\eeq
The fermion fields $\psi_{L/R}$ are represented in terms of chiral boson fields $\phi_{L/R}$ as
\begin{subequations} 
\beq
\psi_L &=& :e^{-\frac{\pi}{2} i \left(\a^L p_L + \b^L p_R \right)} e^{-\sqrt{2} i \phi_L}:
= :e^{-\frac{\pi}{2} i \left(p_L + p_R \right)} e^{-\sqrt{2} i \phi_L}:, \label{n1fermion1}\\
\psi_R &=& :e^{\frac{\pi}{2} i \left(
\a^R p_L + \b^R p_R \right)} e^{\sqrt{2} i \phi_R}:
= :e^{-\frac{\pi}{2} i \left(p_L + p_R \right)} e^{\sqrt{2} i \phi_R} :. \label{n1fermion2}
\eeq
\end{subequations}
It is easy to confirm that the periodic potentials of the boson theory are mapped onto the Dirac mass term and the 
chiral mass term of the fermion theory
\begin{subequations} 
\beq
:e^{\sqrt{2}i\phi}:+ :e^{-\sqrt{2}i\phi}: &=&
\psi^\dag_L \psi_R + \psi^\dag_R \psi_L  =
\barpsi \psi \\
:e^{\sqrt{2}i\phi}: -:e^{-\sqrt{2}i\phi}: &=& \psi^\dag_L \psi_R - \psi^\dag_R \psi_L
= \barpsi \g^5 \psi .
\eeq
\end{subequations}
Here the Dirac gamma matrices are 
\begin{subequations} 
\beq
\gamma^0 &=& \s_1, \quad
\gamma^1 = \s_2, \quad 
\gamma^5 = \s_3 = -i \gamma^0 \gamma^1,\quad
\s_i \s_j = i\e_{ijk} \s_k,   \\
\s_1 &=& \left(\begin{array}{cc}
  0 & 1 \\
  1 & 0 
\end{array}\right),\quad
\s_2 = \left(\begin{array}{cc}
  0 & -i \\
  i & 0 
\end{array}\right),\quad
\s_3 = \left(\begin{array}{cc}
  1 & 0 \\
  0 & -1 
\end{array}\right) .
\eeq
\end{subequations}
We also easily derive the well-known bosonization rules
\begin{subequations}
\beq
\p_\t \phi_{L} &=& - \frac{i}{\sqrt{2}} \psi^\dagger_{L}\psi_{L}, \qquad 
\p_\s \phi_{L} = \frac{1}{\sqrt{2}} \psi^\dagger_{L}\psi_{L},\quad  \label{bosonization1}\\
\p_\t \phi_{R} &=& \frac{i}{\sqrt{2}} \psi^\dagger_{R}\psi_{R}, \qquad 
\p_\s \phi_{R} = \frac{1}{\sqrt{2}} \psi^\dagger_{R}\psi_{R}. \label{bosonization2}
\eeq
\end{subequations}

\subsubsection{Thirring Model}

By the work of Coleman \cite{coleman75} it is well known that the massless Thirring model \cite{Thirring} is equivalent to the 
free boson model. It follows from the above bosonization rules.
\beq
L &=& \frac{1}{2\pi}\left(\bar{\psi} \gamma^\m \p_\m 
\psi + \frac{g}{4\pi} j^\m j_\m\right) \nn\\
&=& \frac{1}{4\pi} (\p \phi)^2 + \frac{g}{4\pi^2} (\p \phi)^2 \\
&=& \frac{1}{4\b^2} (\p \phi)^2,\nn
\eeq
where
\beq
j^\m = \bar \psi \g^\m \psi, ~~~\b^2  = \frac{1}{1+ g/\pi}.
\eeq 
This equivalence can be extended to a more general case of equivalence between the massive Thirring model $L_{Th}$ 
and the sine-Gordon model $L_{SG}$, 
\begin{subequations}
\beq
L_{Th} &=& \frac{1}{2\pi}\left( \barpsi \g \cdot \p \psi + \frac{g}{4\pi} j^\m j_\m \right)+
\frac{m}{2} \barpsi \psi + \frac{m_C}{2i} \barpsi \g^5 \psi, \\
L_{SG} &=& \frac{1}{4 \pi \b^2} \p \phi \p \phi + \frac{m}{2} \left(e^{\sqrt{2}i\phi}+ e^{-\sqrt{2}i\phi}\right)
+ \frac{m_C}{2i} \left(e^{\sqrt{2}i\phi} -e^{-\sqrt{2}i\phi}\right).
\eeq
\end{subequations}

\subsubsection{The Simple Boundary States}
The boundary state is a closed string state which is annihilated by the boundary condition
that would be imposed on an open string embedding function when the open string world-sheet ends on 
the D-brane world-volume. Depending on whether the open string embedding function is longitudinal or
transverse to the brane, the boundary condition is Neumann or Dirichlet, respectively. 
The Neumann and Dirichlet boundary states for the bosonic string obey
 \begin{subequations}
 \begin{eqnarray}
 \phi_L\left|N \right\rangle&=&\phi_R\left|
 N \right\rangle, \label{nn:a}\\
 \phi_L\left| D \right\rangle&=&-\phi_R
 \left|D \right\rangle. \label{dd:b} \end{eqnarray}
 \end{subequations}
The conditions (\ref{nn:a}) and (\ref{dd:b}) are solved by
 \begin{subequations}
 \begin{eqnarray}
 \left| N \right\rangle &=& \sum_{p_L} \prod_{n=1}^\infty \exp\left(
 -a^\dag_n \tilde a^\dag_n\right)\left|
 p_L,-p_L\right\rangle,   \\
 \left| D\right\rangle &=& \sum_{p_L}\prod_{n=1}^\infty\exp \left(
 a^\dag_n\tilde a^\dag_n\right)\left|p_L,p_L\right\rangle.
 \end{eqnarray}
 \end{subequations}
 In the fermion representation with Eqs.~(\ref{n1fermion1}) and
 (\ref{n1fermion2}), the boundary state conditions of boson theory Eqs.~(\ref{nn:a}) and (\ref{dd:b}) are mapped
 onto the following  boundary state conditions of fermion theory 
 \begin{subequations}
  \begin{eqnarray}
  \psi_L~\left|N \right\rangle &=&
  i\psi^{\dagger}_R \left|N \right\rangle~,~~
 \psi_L^{\dagger} \left|N \right\rangle=
  i \psi_R~\left| N\right\rangle, \\
  \psi_L~\left| D\right\rangle &=&
 \psi_R \left| D\right\rangle~,~~
 \psi_L^{\dagger} \left| D\right\rangle=
 - \psi_R^{\dagger}~\left| D\right\rangle.
 \end{eqnarray}
 \end{subequations}
 
With the simplest solution for the Klein factors Eq.(\ref{simple1}), we can also transcribe the Neumann boundary condition and the Dirichlet boundary condition in the boson theory of $N$ species
\begin{subequations} 
\begin{eqnarray}
 \phi^a_L\left|{\bf N} \right\rangle&=&\phi^a_R\left|{\bf N} \right\rangle, \label{LR:a}\\
 \phi^a_L\left| {\bf D} \right\rangle&=&-\phi^a_R
 \left|{\bf D}\right\rangle ,  ~~ a = 1, 2, \dots, N . \label{LR:b}
 \end{eqnarray}
\end{subequations}
into the Neumann boundary condition and 
the Dirichlet boundary condition in the fermion theory of $N$ species as follows
\begin{subequations}
  \begin{eqnarray}
  \psi^a_L~\left|{\bf N} \right\rangle &=&
  i\psi^{a\dagger}_R \left|{\bf N} \right\rangle~,~~
 \psi_L^{a\dagger} \left|{\bf N} \right\rangle=
  i \psi^a_R~\left| {\bf N}\right\rangle, \label{bc1}\\
  \psi^a_L~\left| {\bf D}\right\rangle &=&
 \psi^a_R \left| {\bf D}\right\rangle~,~~
 \psi_L^{a\dagger} \left| {\bf D}\right\rangle=
 - \psi_R^{a\dagger}~\left| {\bf D}\right\rangle, ~~ a = 1, 2, \dots, N . \label{bc2}
 \end{eqnarray}
 \end{subequations}
At a glance the boundary conditions for the Neumann and the Dirichlet states 
Eqs.(\ref{bc1}, \ref{bc2}) appear to be asymmetric. 
By a constant phase shift, 
\begin{subequations}
\beq
\psi^a_L &\rightarrow& e^{-i \frac{\pi}{4}} \psi^a_L, ~~~ \psi^{a\dag}_L \rightarrow e^{i \frac{\pi}{4}} \psi^{a\dag}_L, \\
\psi^a_R &\rightarrow& e^{i \frac{\pi}{4}} \psi^a_R, ~~~ \psi^{a\dag}_R \rightarrow e^{-i \frac{\pi}{4}} \psi^{a\dag}_R,
\eeq
\end{subequations}
they can be made resemble the boundary state conditions of the boson theory
\begin{subequations}
  \begin{eqnarray}
  \psi^a_L~\left|{\bf N} \right\rangle &=&
  i\psi^{a\dagger}_R \left|{\bf N} \right\rangle~,~~
 \psi_L^{a\dagger} \left|{\bf N} \right\rangle=
  i \psi^a_R~\left| {\bf N}\right\rangle, \label{bc3}\\
  \psi^a_L~\left| {\bf D}\right\rangle &=&
 i\psi^a_R \left| {\bf D}\right\rangle~,~~
 \psi_L^{a\dagger} \left| {\bf D}\right\rangle=
 -i \psi_R^{a\dagger}~\left| {\bf D}\right\rangle, ~~ a = 1, 2, \dots, N . \label{bc4}
 \end{eqnarray}
\end{subequations}
These representations of the simple boundary states may be also useful to discuss the resonant multilead 
point-contact tunneling \cite{Nayak} and the multi-channel Kondo problem \cite{Tsvelick85,Ludwig} in the framework of the boundary state 
formulation.


\subsection{Theory with Two Boson Fields: N=2 System}

For the theory with two boson fields or fermion fields with two flavors, the matrices $(\a^{L/R})$ and $(\b^{L/R})$ are given by $2 \times 2$ matrices. The simple solution for this case is
\beq
(\a^L) = -(\a^R) = (\b^L) = -(\b^R) = \begin{pmatrix}
  1 & 2  \\
  0 & 1   
 \end{pmatrix} .
\eeq
We can spell out the explicit expressions of the fermion field operators as follows 
\begin{subequations}
\beq
\psi^1_L &=& e^{-\frac{\pi}{2} i \left(p^1_L +2 p^2_L + p^1_R + 2p^2_R\right)} e^{-\sqrt{2} i \phi^1_L}, \label{n2fermion1}\\
\psi^2_L &=& e^{-\frac{\pi}{2} i \left(p^2_L+ p^2_R\right)} e^{-\sqrt{2} i \phi^2_L}, \label{n2fermion2}\\
\psi^1_R &=& e^{-\frac{\pi}{2} i \left(p^1_L +2 p^2_L + p^1_R + 2p^2_R\right)} e^{\sqrt{2} i \phi^1_R} \label{n2fermion3}\\
\psi^2_R &=& e^{-\frac{\pi}{2} i \left(p^2_L+ p^2_R\right)} e^{\sqrt{2} i \phi^2_R}. \label{n2fermion4}
\eeq
\end{subequations}
This is the same representation of the Klein factors we adopted for the rolling tachyon in 
the previous work \cite{TLee;08}. In the following we discuss applications of the bosonization to the Gross-Neveu model \cite{Gross74,shankar80}, chiral Gross-Neveu model and the Thirring model with two fermion fields. 
Comparing the bosonized actions of the models, we observe some dissimilarities and similarities between the models. The Gross-Neveu model and the chiral Gross-Neveu model have been proposed to 
study the dynamical breaking of the chiral symmetry. Because the models 
share many important features with the quantum chromodynamics in $(3+1)$ dimensions, the asymptotic 
freedom for an example, they are extensively studied in literature. But the studies on the models are 
mainly focused on the perturbative analysis based on the $1/N$ expansion. The Fermi-Bose equivalence 
may enlarge the scope of the analysis, as it enables us to explore non-perturbative domains of the models.  

\subsubsection{Gross-Neveu model with $N=2$}
The Gross-Neveu model with $N$ fermion fields is described by the following
Lagrangian \cite{Gross74}
\beq
L &=& \frac{1}{2\pi}\left[\sum_{a=1}^{N}\barpsi^a \gamma^\m \p_\m \psi^a + \frac{g}{4\pi} 
\left(\sum_{a=1}^{N} \barpsi^a \psi^a\right)^2\right].
\eeq
Making use of the bosonization rule Eqs.(\ref{bosonization1}, \ref{bosonization2}) and the following 
identity which can be proved by the boson field representations of the fermion field operators Eqs.(\ref{n2fermion1},   
\ref{n2fermion2}, \ref{n2fermion3}, \ref{n2fermion4})
\beq \label{bosonizationa}
:(\bar \psi^a \psi^a )^2: &=& -2 \psi^{a\dag}_L \psi^a_L \psi^{a\dag}_R \psi^a_R = - \left(\p \phi^a\right)^2 ,
\eeq
we can map the $N=2$ Gross-Neveu model onto the following Lagrangian for two boson fields \cite{shankar80}
\beq
L &=& \frac{1}{4\pi}\left(1 - \frac{g}{2\pi} \right) \sum_{a=1}^2 \p\phi^a \p\phi^a  \nn\\
&& + 
\frac{g}{4\pi^2} \left(e^{\sqrt{2}i (\phi^1 + \phi^2)}+ e^{-\sqrt{2}i (\phi^1 + \phi^2)}
+e^{\sqrt{2}i (\phi^1 - \phi^2)}+ e^{-\sqrt{2}i (\phi^1 - \phi^2)} \right) .
\eeq
Defining two boson fields $\varphi^a$, $a=1,\, 2$, which are related to two boson fields $\phi^a$, $a=1, 2$ by a
$SO(2)$ transformation 
\beq
\varphi^1 = \frac{1}{\sqrt{2}} \left(\phi^1 + \phi^2 \right), ~~~
\varphi^2 = \frac{1}{\sqrt{2}} \left(\phi^1 - \phi^2 \right), \label{varphi}
\eeq
we rewrite the Lagrangian as 
\beq
L &=& \frac{1}{4\pi}\left(1 - \frac{g}{2\pi} \right) \sum_{a=1}^2 \p\varphi^a \p\varphi^a +\frac{g}{4\pi^2} \left(e^{2i \varphi^1}+ e^{-2i \varphi^1}+ e^{2i \varphi^2}+ e^{-2i \varphi^2} \right) .  
\eeq
Scaling $\varphi^a \rightarrow \varphi^a/\sqrt{2}$, we find that the Gross-Neveu model with $N=2$ is equivalent to 
a direct sum of two sine-Gordon models
\beq
L &=& \sum_{a=1}^2 \left\{\frac{1}{8\pi}\left(1 - \frac{g}{2\pi} \right) \p\varphi^a \p\varphi^a 
+\frac{g}{4\pi^2} \left(e^{i\sqrt{2} \varphi^a}+e^{-i\sqrt{2} \varphi^a}\right) \right\}.
\eeq
This model is critical at $g = - 2\pi$, where it can be shown to be equivalent to a free massive fermion theory with 
two flavors by refermionization 
\beq
L = \frac{1}{2\pi} \sum_{a=1}^2 \bar \psi^a \left( \g^\m \p_\m - m \right) \psi^a , ~~~ m = 1. 
\eeq

\subsubsection{Chiral Gross-Neuveu model with $N=2$}

The chiral Gross-Neveu model with fermion fields of $N$ flavors is described by 
\beq
L &=& \frac{1}{2\pi}\sum_{a=1}^N\barpsi^a \gamma^\m \p_\m \psi^a + \frac{g}{8\pi^2} \left[
\left(\sum_{a=1}^N \barpsi^a \psi^a\right)^2- \left(\sum_{a=1}^N \barpsi^a \g^5 \psi^a\right)^2
\right].
\eeq
Making use of the identities of the bosonization, which follow from the boson representation of the fermion field operators,  
\begin{subequations}
\beq
:\barpsi^a \psi^a: &=& :e^{\sqrt{2}i\phi^a}:+ :e^{-\sqrt{2}i\phi^a}: = 2 :\cos \sqrt{2} \phi^a:, \\
:\barpsi^a \g^5 \psi^a: &=& :e^{\sqrt{2}i\phi^a}:- :e^{-\sqrt{2}i\phi^a}: = 2 i :\sin \sqrt{2} \phi^a:\\
:\left(\barpsi^a \g^5 \psi^a\right)^2: &=& :\left(\psi^{a\dag}_L\psi^a_R - \psi^{a\dag}_R \psi^a_L \right)^2: 
=  :(\p \phi^a)^2: , 
\eeq
\end{subequations}
and Eq.(\ref{bosonizationa}), we can map the $N=2$ chiral Gross-Neveu model onto the following boson model
\beq
L =  \frac{1}{4\pi} \left(1- \frac{g}{\pi} \right)\sum_{a=1}^2 \p\phi^a \p\phi^a + \frac{g}{2\pi^2} 
\left(e^{\sqrt{2} i (\phi^1- \phi^2)}+ e^{-\sqrt{2} i (\phi^1- \phi^2)}\right). 
\eeq
If we take the $SO(2)$ transformation Eq.(\ref{varphi}), as in the case of the Gross-Neveu model, we find 
\beq
L =  \frac{1}{4\pi} \left(1- \frac{g}{\pi} \right)\sum_{a=1}^2 \p\varphi^a \p\varphi^a + \frac{g}{2\pi^2} 
\left(e^{2i \varphi^2}+ e^{-2i \varphi^2}\right). 
\eeq
that the Chiral Gross-Neuveu $N=2$ model is equivalent to a direct sum of a sine-Gordon model and a free boson theory. 
Scaling 
\beq
\varphi^1 \rightarrow \frac{\varphi^1}{\sqrt{1 - \frac{g}{\pi}}}, ~~~ \varphi^2 \rightarrow \frac{\varphi^2}{\sqrt{2}}, 
\eeq
we have 
\beq
L = \frac{1}{4\pi} \p \varphi^1 \p \varphi^1 + 
\frac{1}{8\pi} \left(1-\frac{g}{\pi}\right) \p \varphi^2 \p \varphi^2 + 
\frac{g}{2\pi^2} \left(e^{i \varphi^2}+ e^{-i \varphi^2}\right).
\eeq
This model is critical at $g=-\pi$, where it can be mapped onto a free fermion model by refermionization
\beq
L = \frac{1}{2\pi} \bar \psi^1 \g \cdot \p \psi^1 +  \frac{1}{2\pi} \bar \psi^2 \left(\g \cdot \p - m\right)\psi^2, ~~~ m = 1 . 
\eeq
Here we can see the difference between the Gross-Neveu model and the chiral Gross-Neveu model in their spectrums.

\subsubsection{$SU(2)$ Thirring Model}
The bosonization of the $U(1)$ Thirring model, discussed as an example model with $N=1$, can be generalized to 
the bosonization of $SU(N)\times U(1)$ Thirring model \cite{Banks76}, of which Lagrangian is 
\beq                
L = \frac{1}{2\pi}\left\{\sum_{a=1}^N \barpsi^a \g \cdot \p \psi^a + 
\frac{g_{U(1)}}{4\pi} J^\m J_\m + \frac{g_{SU(N)}}{4\pi}\sum_i J^{\m (i)}J_\m{}^{(i)} \right\}, 
\eeq 
where $J^\m$ and $J^{\m(i)}$ are $U(1)$ current and $SU(N)$ current respectively,
\beq
J^\m = \sum_{a=1}^N \barpsi^a \g^\m \psi^a, \quad 
J^{\m(i)} = \sum_{a,\,b =1}^N \barpsi^a \g^\m \frac{\lambda^{(i)}_{ab}}{2} \psi^b, ~~~ i = 1, 2, \dots, N,
\eeq
and $\l^{(i)}_{ab}$, $i = 1, 2, \dots, N$, are generators of $SU(N)$ group. The case of $N=2$, {\it i.e.}, $SU(2)$ Thirring model has been discussed in the literature \cite{Banks76} in the context of bosonization. But an explicit representation of the Klein factors has not been given and the bosonized action contains the products of Klein factors. Here we discuss bosonization of the  $SU(2)$ Thirring model explicitly, using the Klein factors given by
Eqs.(\ref{n2fermion1}, \ref{n2fermion2}, \ref{n2fermion3}, \ref{n2fermion4}). We shall see that the bosonized
action of the model does not contain any non-trivial Klein factor. 

The bosonized form of $U(1)$ current is found as 
\beq
J^0 = \sqrt{2}\sum_{a=1}^2 \p_\s \phi^a, \quad J^1 =- \sqrt{2}\sum_{a=1}^2 \p_\t \phi^a .
\eeq
Thus, we may rewrite the $U(1)$ current term as 
\beq
J^\m J_\m &=& (J^0)^2+(J^1)^2 = 2 \left(\sum_{a=1}^2 \p_\s \phi^a\right)^2 + 
2 \left(\sum_{a=1}^2 \p_\t \phi^a\right)^2 \nn\\
&=& 2 \sum_{a, b =1}^2 \p \phi^a \p \phi^b . \label{U1}
\eeq
In order to find the bosonized form of the $SU(2)$ Thirring interaction term, we need to do some algebra.
Choosing $\l^{(i)}_{ab} = \s^i_{ab}$, $i = 1, 2, 3$ for $SU(2)$, we may write the $SU(2)$ currents 
as
\begin{subequations}
\beq
J^{\mu (1)} &=& \half \barpsi^a \g^\m (\s^1)_{ab} \psi^b = \half \left(\barpsi^2 \g^\m \psi^1+ 
\barpsi^1 \g^\m \psi^2 \right), \\
J^{\mu (2)} &=& \half \barpsi^a \g^\m (\s^2)_{ab} \psi^b = \frac{i}{2} \left(\barpsi^2 \g^\m \psi^1-
\barpsi^1 \g^\m \psi^2 \right),\\
J^{\mu (3)} &=& \half \barpsi^a \g^\m (\s^3)_{ab} \psi^b = \frac{1}{2} \left(\barpsi^1 \g^\m \psi^1-
\barpsi^2 \g^\m \psi^2 \right) .
\eeq
\end{subequations}
Although the $SU(2)$ currents may not be expressed entirely in terms of the boson fields, the 
$SU(2)$ Thirring interaction term may be bosonized to be entirely rewritten in terms of boson fields only.
The $SU(2)$ Thirring interaction term may be written as 
\beq
\sum_{\m,\, i} J^{\m (i)} J_\m{}^{(i)} &=& \frac{1}{4} \sum_{\m, i, a, b, c, d} (\barpsi \g^\m)_a \s^i_{ab} \psi_b 
(\barpsi \g^\m)_c \s^i_{cd} \psi_d \nn\\
&=& \frac{1}{4} \sum_{\m, a, b} \left[ 
2 (\barpsi \g^\m)_a \psi_b  (\barpsi \g^\m)_b \psi_a - 
(\barpsi \g^\m)_a \psi_a  (\barpsi \g^\m)_b \psi_b \right],
\eeq
where the completeness relation of the Pauli matrices is used
\beq
\sum_{i=1}^3 \s^i_{ab} \,\s^i_{cd} = 2 \d_{ad} \d_{bc} - \d_{ab} \d_{cd}.
\eeq
Rewriting the $SU(2)$ Thirring interaction term explicitly in terms of the $SU(2)$ components of the 
fermi fields, we have
\beq
\sum_{\m, i} J^{\m (i)} J_\m{}^{(i)}
&=& \sum_{a,\,b} \left(\psi^{a\dagger}_L \psi^b_L \psi^{b\dagger}_R \psi^a_R+ 
\psi^{a\dagger}_R \psi^b_R \psi^{b\dagger}_L \psi^a_L\right) 
- \frac{1}{4} \left(\sum_a \barpsi^a \g^\m \psi^a\right)^2 \nn\\
&=& 2\sum_{a}\psi^{a\dagger}_L \psi^a_L \psi^{a\dagger}_R \psi^a_R - \sum_{a \not=b} \left(U_a U^\dag_b + U_b U^\dag_a \right) - \frac{1}{4} J^\m J_\m \nn\\
&=& \sum_a (\p \phi^a)^2 
-\sum_{a\not=b} \left[ e^{i\sqrt{2}(\phi^a-\phi^b)} + e^{-i\sqrt{2}(\phi^a-\phi^b)}\right]
- \frac{1}{4} J^\m J_\m. 
\eeq
Here it should be noted that with the representation the fermion field operators Eqs.(\ref{n2fermion1}, \ref{n2fermion2}, \ref{n2fermion3}, \ref{n2fermion4}), the Klein factors do not explicitly enter the bosonized $SU(2)$ Thirring interaction term. 

Making use of the bosonized form of the $U(1)$ and $SU(2)$ Thirring interaction terms, we are able to rewrite the Lagrangian of $SU(2) \times U(1)$ Thirring model entirely 
in terms of the boson fields
\beq
L &=& \frac{1}{4\pi} \left(1 + \frac{g_{SU(2)}}{2\pi} \right) \sum_{a=1}^2 \left(\p \phi^a \right)^2 +
\frac{1}{4\pi^2} \left(g_{U(1)} - \frac{g_{SU(2)}}{4} \right) \sum_{a,b=1}^2 \p \phi^a \p \phi^b \nn\\
&&- \frac{g_{SU(2)}}{4 \pi^2}
\left(e^{i\sqrt{2}(\phi^1 - \phi^2)}+ e^{-i\sqrt{2}(\phi^1 - \phi^2)} \right) .
\eeq
The boson Lagrangian may be simplified if rewritten by the two boson fields $\varphi^a$ $a=1, 2$, 
defined by Eq.(\ref{varphi}) as in the case of the $N=2$ Gross-Neuveu model
\beq
L = \frac{1}{4\pi} \left(1+ \frac{2 g_{U(1)}}{\pi} \right) \p \varphi^1 \p \varphi^1 + 
\frac{1}{4\pi} \left(1+ \frac{ g_{SU(2)}}{2\pi} \right) \p \varphi^2 \p \varphi^2 - \frac{g_{SU(2)}}{4\pi^2}
\left(e^{2i \varphi^2}+ e^{-2i \varphi^2} \right) .
\eeq
Scaling two boson fields $\varphi^1$ and $\varphi^2$ as 
\beq
\varphi^1 \rightarrow \frac{\varphi^1}{\sqrt{1+ \frac{2g_{U(1)}}{\pi}}}, ~~~ \varphi^2 \rightarrow \frac{\varphi^2}{\sqrt{2}},
\eeq
we find that the $SU(2)\times U(1)$ Thirring model is equivalent to a direct sum of a free boson model and a sine-Gordon model
\beq
L = \frac{1}{4\pi} \p \varphi^1 \p \varphi^1+ \frac{1}{8\pi} \left(1+ \frac{ g_{SU(2)}}{2\pi} \right) \p \varphi^2 \p \varphi^2 - \frac{g_{SU(2)}}{4\pi^2}\left(e^{i\sqrt{2} \varphi^2}+ e^{-i\sqrt{2} \varphi^2} \right).
\eeq
This model is critical at the point where $g_{SU(2)} = 2\pi$. 
Refermionizing the model at this point brings us to a free fermion model which contains 
a massless fermion field and a massive fermion field
\beq
L &=& \frac{1}{2\pi} \bar\psi^1 \g \cdot \p \psi^1 + 
\frac{1}{2\pi} \bar\psi^2 \left(\g \cdot \p -m \right) \psi^2, ~~~ m =1.
\eeq
It is interesting to note that the $SU(2)$ Thirring model is equivalent to the $N=2$ chiral Gross-Neveu model. 
The spectra of both models at the critical point are in exact agreement. 


\subsection{Theory with Three Boson Fields: $N=3$ System}

For the theory with three boson fields or fermion fields with three flavors, the matrices $(\a^{L/R})$ and $(\b^{L/R})$ are given by $3 \times 3$ matrices. The simple solution for this case is
\beq
(\a^L) = -(\a^R) = (\b^L) = -(\b^R) = \begin{pmatrix}
  1 & 2 & 2 \\
  0 & 1 & 2 \\
  0 & 0 & 1 
 \end{pmatrix} .
\eeq
The explicit expressions of the fermion field operators are written as follows 
\begin{subequations}
\beq
\psi^1_L &=& e^{-\frac{\pi}{2} i \left(p^1_L +2 p^2_L + 2p^3_L + p^1_R + 2p^2_R+ 2p^3_R\right)} e^{-\sqrt{2} i \phi^1_L}, \label{n3fermion1}\\
\psi^2_L &=& e^{-\frac{\pi}{2} i \left(p^2_L+ 2 p^3_L + p^2_R + 2p^3_R \right)} e^{-\sqrt{2} i \phi^2_L}, \label{n3fermion2}\\
\psi^3_L &=& e^{-\frac{\pi}{2} i \left(p^3_L + p^3_R  \right)} e^{-\sqrt{2} i \phi^3_L}, \label{n3fermion3}\\
\psi^1_R &=& e^{-\frac{\pi}{2} i \left(p^1_L +2 p^2_L+ 2p^3_L + p^1_R + 2p^2_R+ 2p^3_R\right)} e^{\sqrt{2} i \phi^1_R}, \label{n3fermion4}\\
\psi^2_R &=& e^{-\frac{\pi}{2} i \left(p^2_L+2 p^3_L+ p^2_R+ 2p^3_R\right)} e^{\sqrt{2} i \phi^2_R}, \label{n3fermion5}\\
\psi^3_R &=& e^{-\frac{\pi}{2} i \left(p^3_L + p^3_R  \right)} e^{\sqrt{2} i \phi^3_R}. \label{n3fermion3}
\eeq
\end{subequations}
As example models for the theory with three boson fields, we choose to discuss the quantum Brownian motion on a 
triangular lattice and the chiral Gross-Neveu model with $N=3$.

\subsubsection{Quantum Brownian Motion on a Triangular Lattice}

The application of the Fermi-Bose equivalence has been studied in ref.\cite{Lee2009q} in some detail. Here 
we review the subject briefly for the purpose of the purpose of comparison.
The Euclidean action for the quantum Brownian motion (QBM) is given as follows  
\beq \label{action1}
S_{QBM} &=& \frac{\eta}{4\pi} \int^{\beta}_{0} dt dt^\prime 
\frac{\left({\bf X}(t) - {\bf X}(t^\prime)\right)^2}{(t-t^\prime)^2} 
+ \frac{M}{2} \int^{\beta}_{0} dt \dot {\bf X}^2 + V_0\int^{\beta}_{0} dt \sum_{i=1}^3 \cos \left(2\pi{\bf k}_i \cdot {\bf X}\right) 
\eeq
where $\beta=1/T$ and     
\beq \label{lattice}
{\bf k}_1 = (\frac{1}{2},\frac{\sqrt{3}}{2}), ~~~
{\bf k}_2 = (\frac{1}{2},-\frac{\sqrt{3}}{2}), ~~~
{\bf k}_3 = \left(-1,0 \right).
\eeq
The first non-local action depicts the frictional force due to the coupling of the particles to a bath or an environment which consists of an infinite set of Harmonic oscillators \cite{caldeira83ann,caldeira83phy} 
and the third term is the periodic potential on the triangular lattice. Since the triangular lattice spans a two
dimensional plane, the model contains only two boson fields initially. But in order to fermionize the model,
it is necessary to introduce an auxiliary boson field $X^3$. Trading the non-local term on one dimension
with the Polyakov local action of string theory on two dimensions, we may rewrite the QBM action as 
\beq
S = \frac{\a}{4\pi} \int d\t d\s \p_\a \phi^a \p^\a \phi^a + \frac{V}{2}
\int d\s \sum_{a=1}^3 \left(e^{\frac{i}{\sqrt{2}}\left(\phi^a-\phi^{a+1}\right)}
+e^{-\frac{i}{\sqrt{2}}\left(\phi^a-\phi^{a+1}\right)} \right)
\eeq
where $\phi^{a+3} = \phi^a$. The three boson fields $\phi^a,~ a =1 ,2 ,3$ are related 
to $X^a, ~a=1, 2, 3$ by an $O(3)$ rotation
\begin{subequations}
\beq
\phi^1 &=& \frac{1}{\sqrt{2}} X^1 
+ \frac{1}{\sqrt{6}} X^2 + \frac{1}{\sqrt{3}} X^3, \\
\phi^2 &=& -\frac{1}{\sqrt{2}} X^1 
+ \frac{1}{\sqrt{6}} X^2 + \frac{1}{\sqrt{3}} X^3, \\
\phi^3 &=& -\sqrt{\frac{2}{3}} \,\,X^2 + \frac{1}{\sqrt{3}} X^3. 
\eeq
\end{subequations}

If the periodic boundary interaction is absent, the boundary conditions for $\phi^a$,
$a = 1, 2, 3$,  
would be Neumann: $ \left(\phi^a_L - \phi^a_R\right)\vert_{\s = 0} = 0$. Thus, the boundary state for QBM may be written formally as 
\beq
|B_{QBM}\rangle = :\exp\left[-\frac{V}{2}
\int d\s \sum_{a=1}^3 \left(e^{\frac{i}{\sqrt{2}}\left(\phi^a-\phi^{a+1}\right)}
+e^{-\frac{i}{\sqrt{2}}\left(\phi^a-\phi^{a+1}\right)} \right)\right]:|{\bf N} \rangle
\eeq
where
$\left(\phi^a_L- \phi^a_R\right)|{\bf N}\rangle = 0$, $a = 1, 2, 3$.

With the Neumann condition, we may write the boundary state for QBM at the critical point in fermion theory as 
\beq \label{boundstate}
|B_{QBM}\rangle &=& :\exp\left[ 
\frac{V}{2} \int d\s \sum_{a=1}^3 \left(\psi^{a\dagger}_L \psi^{a+1}_L - \psi^{a+1 \dagger}_L \psi^a_L\right) \right]:|{\bf N}\rangle, \nn\\
&=& :\exp\left[\frac{V}{2} \int d\s \sum_{a=1}^3 \left(\psi^{a+1 \dagger}_R \psi^a_R-
\psi^{a\dagger}_R \psi^{a+1}_R \right) \right]: |{\bf N}\rangle, 
\eeq
where $\psi^{a+3}_{L/R} = \psi^a_{L/R}$.
From the boundary state Eq.(\ref{boundstate}), 
rewritten in terms of the fermion fields, we may deduce the fermionized action for the 
QBM as follows 
\beq
S_{QBM} = \frac{1}{2\pi} \int d\t d\s \sum_{a=1}^3 \left(\bar\psi^a \gamma^\mu \p_\mu \psi^a+ \frac{g}{4\pi} j^{\mu a} j_\m^a \right) + \frac{V}{4} \int d\s \sum_{a=1}^3
\left(\bar\psi^a \gamma^1 \psi^{a+1} - \bar \psi^{a+1} \gamma^1 \psi^a \right)
\eeq
where $g = \pi(\a-1)$ and $j^{\mu a} = \bar\psi^a \gamma^\mu \psi^a$.
Thus, the QBM model is equivalent to a generalized Thirring model with
boundary terms, which are quadratic in fermion fields.

\subsubsection{Chiral Gross-Neveu Model with $N=3$}

Applying the Fermi-Bose equivalence with the constructed Klein factors to the chiral Gross-Neveu model
with $N=3$, 
\beq
L &=& \frac{1}{2\pi}\sum_{a=1}^N\barpsi^a \gamma^\m \p_\m \psi^a + \frac{g}{8\pi^2} \left[
\left(\sum_{a=1}^3 \barpsi^a \psi^a\right)^2- \left(\sum_{a=1}^3 \barpsi^a \g^5 \psi^a\right)^2
\right],
\eeq
we find that its bosonized Lagrangian is given as
\beq
L= \frac{1}{4\pi} \left(1- \frac{g}{\pi} \right) \sum_{a=1}^3 \p\phi^a \p\phi^a +  \frac{g}{2\pi^2} \sum_{a=1}^3
\left(e^{\sqrt{2} i (\phi^a- \phi^{a+1})}+ e^{-\sqrt{2} i (\phi^a- \phi^{a+1})}\right)
\eeq
where $\phi^{3+1} = \phi^1$. 
Introducing three boson fields $\varphi^a$, $a=1, 2, 3$, which are related to the three boson fields 
$\phi^a$, $a=1, 2, 3$ by scaling and $SO(3)$ rotation 
\begin{subequations}
\beq
\varphi^1 &=& \sqrt{2} \phi^1 - \sqrt{2} \phi^2, \\
\varphi^2 &=& \sqrt{\frac{2}{3}} \phi^1 + \sqrt{\frac{2}{3}} \phi^2 -2 \sqrt{\frac{2}{3}} \phi^3, \\
\varphi^3 &=&  \frac{2}{\sqrt{3}} \phi^1 + \frac{2}{\sqrt{3}} \phi^2 +\frac{2}{\sqrt{3}} \phi^3 ,  
\eeq
\end{subequations}
we may rewrite the bosonized Lagrangian of the chiral Gross-Neveu model with $N=3$ as 
\beq \label{bosonizedgs}
L = \frac{1}{16\pi} \left( 1 - \frac{g}{\pi}\right) \sum_{a=1}^3 \p \varphi^a  \p \varphi^a + \frac{g}{2\pi^2} 
\sum_{i=1}^3 \cos \left( {\bf k}_i \cdot \bolvarphi \right) .
\eeq
Here ${\bf k}_i$ $i= 1, 2, 3$, are the three vectors Eq.(\ref{lattice}), which spans the triangular lattice and $\bolvarphi$ in the periodic potential is a two dimensional vector $(\varphi^1, \varphi^2)$. 
Since the boson field $\varphi^3$ does not enter the periodic potential, it may trivially integrated out. 
Thus, the chiral Gross-Neveu model with $N=3$ is equivalent to the sine-Gordon model on a two dimensional 
triangular lattice.

\section{Conclusions}

The Klein factors are important ingredients of the Fermi-Bose equivalence, which has been an essential 
tool to analyze a wide variety of $(1+1)$ dimensional models in condensed matter physics, quantum field theory and string theory. However, the conventional representations of the Klein factors have room for improvement. 
They may enter the action and the boundary state conditions explicitly if we transcribe a boson theory 
into its corresponding fermion theory and vice versa, using the Fermi-Bose equivalence. 
It is certainly an undesirable feature of the conventional representations of the Klein factors. In this paper 
we completed the construction of a new representation of the Klein factors, which has been initiated in 
recent works \cite{Lee:2005ge, Hassel, Tlee:06, TLee;08, Lee2009q}. 
The new representation of the Klein factors is a generalization of the kink or soliton
operator in the anisotropic Heisenberg model, of which role is to change the statistics of the spin-1/2 operator. 
The new representation of the Klein factors resolves the problems commonly shared 
by the conventional representations as they do not make an explicit appearance in the action and the boundary 
state conditions.  

We wrote down the most general form of the Klein factors for the theory with $N$ boson fields, which has $4N^2$ parameters. Even if we impose the condition that the fermion field operators satisfy the anti-commutation relations, $(2N^2+1)$ parameters remain unfixed. So we are able to impose further constraints such that 
the Klein factors do not appear in the actions and the boundary states when we map the $(1+1)$ dimensional theories by applying the Fermi-Bose equivalence. By an explicit construction we have shown 
that these conditions are consistent and there exist solutions satisfying all these conditions. 
A simple solution has been presented 
as an explicit example and applied to some well-known $(1+1)$ dimensional theories. 

We applied the Fermi-Bose equivalence to the simple boundary states of the theory with arbitrary number
of boson fields. Then we chose Gross-Neveu model with $N=2$, 
chiral Gross-Neveu models with $N=2$ and $3$, and $SU(2)$ Thirring model to apply the Fermi-Bose equivalence,
using the newly constructed representation of Klein factors. The Gross-Neveu model with $N=2$ was shown to be 
equivalent to a direct sum of two sine-Gordon models while the chiral Gross-Neveu model with $N=2$ is equivalent to a direct sum of a sine-Gordon model and a free boson model. An interesting observation is that the $N=2$ chiral 
Gross-Neveu model is equivalent to the $SU(2)$ Thirring model. As example models for the theory with three 
boson fields, we discussed a model for the quantum Brownian motion (QBM) on a triangular lattice and the $N=3$ chiral Gross-Neveu model. The model for QBM is defined initially with two boson fields. But to fermionize the 
model an auxiliary boson field is introduced additionally. Thus, in order to apply the Fermi-Bose 
equivalence, we need to redefine the model with three boson fields. Applying the Fermi-Bose equivalence, 
we obtained a Thirring model type action with a fermion bilinear interaction on the boundary. Since a similar 
action arises in the single-channel spinless Tomonaga-Luttinger liquid model for the junction of three
quantum wires \cite{oshikawa2005}, the new representation of the Klein factors may be useful to study the 
multi-junctions of quantum wires. 

A more interesting example model may be the chiral Gross-Neveu model with $N=3$. If the model is bosonized, 
the phase interaction, which appears on the spatial boundary in the case of QBM model, 
emerges in the bulk action of the model. Then by taking a $SO(3)$ rotation of the three boson fields, 
the bosonized action may be rewritten as a sine-Gordon model on a two dimensional triangular lattice. This action 
can be identified as the effective field theory action for the long wavelength fluctuations of the fully packed loop (FPL) model in statistical physics \cite{Reshetikhin91,Blote94,Kondev}. 
The chiral Gross-Neveu model with $N=3$ corresponds to the 
FPL model with loop fugacity $n=2$ model, which undergoes a Kosterlitz-Thouless type transition \cite{Reshetikhin91,Kondev94} into a long-range ordered state. Applications of the 
Fermi-Bose equivalence to the FPL models and the 
reinterpretation of the phase transitions of the FPL model in the context of the chiral Gross-Neveu model 
could be excellent subjects, which deserve further study.


\vskip 1cm

\begin{acknowledgments}
This work was supported by Kangwon National University Research Grant 2013. 
\end{acknowledgments}


%

%





\end{document}